\documentclass[conference]{IEEEtran}
\IEEEoverridecommandlockouts
\usepackage{cite}
\usepackage{amsmath,amssymb,amsfonts}
\usepackage{algorithmic}
\usepackage{graphicx}
\usepackage{tikz}
\usepackage{pgfplots}
\usepgfplotslibrary{fillbetween}
\usepackage{multirow}
\usetikzlibrary{arrows,shapes}
\usetikzlibrary{datavisualization}
\tikzset{vertex/.style={circle,draw,minimum size=1em}}
\tikzset{edge/.style={draw}}

\usepackage{hyperref}

\pgfplotsset{compat=newest}
\definecolor{tb_color_1}{RGB}{245,124,0}
\definecolor{tb_color_2}{RGB}{0,167,247}

\pgfplotscreateplotcyclelist{tb}{
semithick,tb_color_1\\%
semithick,tb_color_2\\%
}

\usepackage{textcomp}
\usepackage{xcolor}
\def\BibTeX{{\rm B\kern-.05em{\sc i\kern-.025em b}\kern-.08em
    T\kern-.1667em\lower.7ex\hbox{E}\kern-.125emX}}

\newcommand{\quotes}[1]{``#1''}

\begin{document}

\title{Symbolic Music Genre Transfer with CycleGAN\\
}

\author{Gino Brunner, Yuyi Wang, Roger Wattenhofer and Sumu Zhao* \\
% Distributed Computing Group \\
 Department of Information Technology and Electrical Engineering\\
        ETH Z{\"u}rich\\
        Switzerland\\
        {\tt\small brunnegi,yuwang,wattenhofer,suzhao@ethz.ch}
\thanks{*
\bf{Authors are listed in alphabetical order.}}% <-this % stops a space
}

\maketitle

\begin{abstract}
Deep generative models such as Variational Autoencoders (VAEs) and Generative Adversarial Networks (GANs) have recently been applied to style and domain transfer for images, and in the case of VAEs, music.
GAN-based models employing several generators and some form of cycle consistency loss have been among the most successful for image domain transfer. In this paper we apply such a model to symbolic music and show the feasibility of our approach for music genre transfer. Evaluations using separate genre classifiers show that the style transfer works well. In order to improve the fidelity of the transformed music, we add additional discriminators that cause the generators to keep the structure of the original music mostly intact, while still achieving strong genre transfer. Visual and audible results further show the potential of our approach. To the best of our knowledge, this paper represents the first application of GANs to symbolic music domain transfer. 
\end{abstract}


\section{Introduction}

Style and domain transfer using neural networks have become exciting machine learning showcases. Most prior work has focused on the image domain, and has enabled us, for example, to take photographs and have them rendered in the style of a certain painter~\cite{DBLP:conf/cvpr/GatysEB16}, or change an image taken during summer to look like it were captured in winter~\cite{DBLP:conf/iccv/ZhuPIE17}. 
% Possible applications of style transfers are only limited by our imagination. 
Domain transfers are interesting, because they require the development of novel representation learning techniques that will carry over to other areas in Deep Learning research. In order for domain transfers to work, the neural network models must have a \quotes{deep} understanding of the underlying domain. This requires the extraction of salient features from complex data such as images, natural language, or music. Deep generative models like Variational Autoencoders~\cite{DBLP:journals/corr/KingmaW13} (VAE) and Generative Adversarial Networks~\cite{goodfellow2014generative} (GAN) seem well suited for this task, as they attempt to learn the true underlying data generating distribution. Thus, neural style transfer using deep generative models is a highly relevant part of deep representation learning research~\cite{DBLP:journals/pami/BengioCV13}.

Domain transfer for music has many possible real world applications. For instance, professional musicians often create cover songs, i.e., new interpretations of a song from another musician. If both musicians roughly belong to the same genre, a slight change in instrumentation coupled with the unique characteristics of the cover artist's voice could already be enough to make the cover song worth listening to. However, there are many cases were the original and cover artists come from completely different styles. 
In such cases, the transformations necessary to make the cover song pleasing to listen to are far more elaborate. One can only imagine the amount of effort that goes into arranging an entire symphony based on a comparatively simple rock song~\cite{bohemianrphasclassic}. Domain transfer systems could significantly accelerate this process, or even automate it completely, which would let us enjoy music that generally has not been feasible to create on a large scale.

The terms style and domain transfer have often been used interchangeably in the literature. As there are no standard definitions or distinctions of the two terms, which can lead to some confusion, we will discuss them briefly at this point.
The term style transfer in the context of neural networks was introduced by Gatys et al.~\cite{DBLP:conf/cvpr/GatysEB16} and usually refers to preserving explicit content features of an image and applying to it explicit style features of another image. The explicit style and content features are, e.g., extracted from a pre-trained CNN. Thus, style transfer enables the merging of two images while allowing the control over how much style and content each of the images contributes. The concept of domain transfer is more general, as it aims to learn a mapping between entire domains of, e.g. images. For instance, domain transfer allows to take any input from domain $A$ and change it such that it looks like it belongs to domain $B$, where $A$ and $B$ could be summer and winter, or Jazz and Classic. There must not necessarily be an explicit constraint that preserves \quotes{content}, but regularization techniques such as applying a cycle consistency loss
as in \cite{DBLP:conf/iccv/ZhuPIE17} encourage the preservation of overall content, and help the network to only change what is necessary to perform the domain transfer. However, we think that the idea of style transfer is more general than simply using the \quotes{style} features extracted from a CNN, and that it depends on the definition of style.
For example, if we want to transfer music from one genre to another, it could either be called style transfer (if style is defined as the genre) and/or domain transfer, where music genres $A$ and $B$ are seen as two domains. Nevertheless, for the sake of consistency, we will use the terms domain or genre transfer, but still occasionally use the term style transfer when referring to prior work.

In this paper we consider the task of transferring a piece of music from a source to a target genre. The transfer should be clearly noticeable, while retaining enough of the original melody and structure such that the source piece is still recognizable. To that end we adapt CycleGAN~\cite{DBLP:conf/iccv/ZhuPIE17}, a successful neural domain transfer architecture for images, to perform genre transfer on symbolic music. We show that our model can transform polyphonic music pieces from a source to a target genre, e.g., from Jazz to Classic, by only changing note pitches. 
We introduce additional discriminators to balance the strength of the domain transfer against retaining the original input's content. 
We use separate genre classifiers to quantify the effect of the genre transfer. Provided audio samples show that the genre transfer cannot only be picked up by a neural network classifier, but can indeed be heard by humans. Additionally, the polyphonic music generated by our model sounds pleasing and harmonic, with relatively few dissonant notes or rhythmic stumbles. To the best of our knowledge, we present the first successful attempt at domain transfer for symbolic music with GANs. 
In order to facilitate future research we provide our code and training data.\footnote{Repository:\\ \url{https://github.com/sumuzhao/CycleGAN-Music-Style-Transfer}}

\section{Related Work}
\label{relatedWork}

Gatys et al.~\cite{DBLP:conf/cvpr/GatysEB16} introduce the concept of neural style transfer and show that pre-trained CNNs can be used to merge the style and content of two images. Approaches such as CycleGAN~\cite{DBLP:conf/iccv/ZhuPIE17} do not require the extraction of explicit style and content features, but instead uses a pair of generators to transform data from a domain A to another domain B. The nature of the two domains implicitly specifies the kinds of features that will be extracted. For example, if domain A contains photographs and domain B contains paintings, then CycleGAN should learn to transfer any painting into a photograph and vice versa. 
% The cycle consistency loss of CycleGAN makes sure that \quotes{content} is preserved when transferring between domains
We use the same structure as CycleGAN and apply it to music in the MIDI format. The general idea of CycleGAN has been further developed and improved. A few notable examples include CycleGAN-VC~\cite{DBLP:journals/corr/abs-1711-11293}, StarGAN~\cite{DBLP:journals/corr/abs-1711-09020}, CoGAN~\cite{DBLP:conf/nips/LiuT16} and DualGAN~\cite{DBLP:conf/iccv/YiZTG17}. In the future we plan on using a more complex architecture and incorporate improvements from these works, but in this paper we focus on showing the feasibility of a CycleGAN approach to domain transfer for symbolic music. 
% Since then, more powerful approaches have been developed \cite{DBLP:conf/nips/LiFYWLY17,DBLP:conf/iccv/ZhuPIE17}; these allow, for example, to render an image taken in summer to look like it was shot in winter. 

Existing work on music style transfer includes Malik et al.~\cite{DBLP:journals/corr/abs-1708-03535}, who introduce a model that learns to play music in the style of a human musician. Their model adds velocities to \quotes{flat} MIDI files which results in more realistic sounding music. While their model can indeed 
play music in a more human-like manner, it can only change note velocities, and does not learn the characteristics of different musical styles/genres.
Brunner et al.~\cite{midivaeismir2018} create MIDI-VAE, a multi-task Variational Autoencoder model with a shared latent space that is capable of changing the style of complete compositions from, e.g., Classic to Jazz. In addition to note pitches, MIDI-VAE also models most other aspects of music contained in MIDI files, i.e., velocities, note durations and instrumentation. In contrast to MIDI-VAE, we do not limit the number of simultaneously played notes, which leads to richer sounding music. Furthermore, when only considering the note pitches, our method achieves a more convincing style transfer.
For raw audio, Van den Oord et al.~\cite{DBLP:conf/nips/OordVK17} introduce a VAE model with discrete latent space that is able to perform speaker voice transfer. Mor et al.~\cite{DBLP:journals/corr/abs-1805-07848} develop a system based on WaveNet~\cite{DBLP:conf/ssw/OordDZSVGKSK16} autoencoders 
that is capable of translating raw music between instruments, genres and styles. Their system even enables the synthesis of music from whistling.

The focus of this paper lies on musical genre transfer. However, genre transfer can only be successful if the resulting music sounds pleasant. Therefore we will briefly cover important work in the field of automatic music generation without direct application to style or domain transfer. 
% \nb{@yuyi: use present or past tense here?}
Much of the existing work uses standard Recurrent Neural Networks (RNN) (\hspace{-1sp}\cite{todd1989connectionist,DBLP:journals/connection/Mozer94}) or long short-term memory networks~\cite{DBLP:journals/neco/HochreiterS97} (\hspace{-1sp}\cite{eck2002first,DBLP:conf/icml/Boulanger-LewandowskiBV12,brunner2017jambot,DBLP:conf/icml/HadjeresPN17,DBLP:journals/corr/ChuUF16}) to model music. More recently, CNNs have also been successfully applied, sometimes in combination with RNNs (\hspace{-1sp}\cite{DBLP:conf/evoW/Johnson17,chuan2018modeling}).
Generative models such as the Variational Autoencoder (VAE) and Generative Adversarial Networks (GANs) have been increasingly successful at generating music.
% The introduction of new generative models, namely the Variational Autoencoder (VAE) and Generative Adversarial Networks (GANs), brought about countless impressive works in the visual domain. Progress has been slower in sequence tasks, but the pace has recently picked up, and some of the best systems are now based on generative models. 
% Google Brain's Magenta project introduced a range of new models for the generation of symbolic music. One of the most recent additions is 
Roberts et al. introduce MusicVAE~\cite{roberts2017hierarchical}, a hierarchical VAE model that can capture long-term structure in polyphonic music and exhibits high interpolation and reconstruction performance. 
% They use fixed instrument assignments and treat drums separately. Our model is also based on VAE, but it can learn arbitrary instrument mappings that do not need to be defined beforehand. 
GANs, while very powerful, are notoriously difficult to train and have generally not been applied to sequential data. However, Mogren~\cite{DBLP:journals/corr/Mogren16}, Yang et al.~\cite{DBLP:conf/ismir/YangCY17} and Dong et al.~\cite{DBLP:conf/aaai/DongHYY18} have recently shown the efficacy of CNN-based GANs for music composition. We use CNN-based GANs to model music and perform domain transfer.
Yu et al.~\cite{DBLP:conf/aaai/YuZWY17} were the first to successfully apply RNN-based GANs to music by incorporating reinforcement learning techniques.
For a more comprehensive overview of automatic music generation, we refer the interested reader to the following surveys: \cite{DBLP:journals/jair/FernandezV13,briot2017deep,DBLP:journals/csur/HerremansCC17}.
% \nb{the order of related work is a bit confusing. i would always go from "closest to farthest" or "farthest to closest" "far-close-far-close"}

% Approaches to generate raw audio waves have also been proposed, though it is usually considered much more difficult, because the domain of raw audio is very high dimensional. Based on CNNs, Van den Oord et al.~\cite{DBLP:conf/ssw/OordDZSVGKSK16} introduced WaveNet, for the conditional generation of speech. 
% They also showed that it can be used to generate pleasing sounding piano music. 
% More recently, Engel et al.~\cite{DBLP:conf/icml/EngelRRDNES17} incorporated WaveNet into an Autoencoder structure to generate musical notes and different instrument sounds. 
% Mehri et al.~\cite{DBLP:journals/corr/MehriKGKJSCB16} developed SampleRNN, for unconditional generation of raw audio. 

% Thus most existing work on music generation uses symbolic music representations (see e.g., \cite{DBLP:journals/corr/abs-1708-03535,DBLP:conf/icmc/Cope87,todd1989connectionist,DBLP:journals/connection/Mozer94,DBLP:conf/icml/Boulanger-LewandowskiBV12,DBLP:journals/corr/ChuUF16,brunner2017jambot,DBLP:conf/icml/HadjeresPN17,DBLP:conf/evoW/Johnson17,chuan2018modeling,roberts2017hierarchical,DBLP:journals/corr/Mogren16,DBLP:conf/ismir/YangCY17,DBLP:journals/corr/abs-1709-06298,DBLP:conf/aaai/YuZWY17}). 

\section{Model Architecture}\label{sec:architecture}

\begin{figure*}[!t] 
\centering
\includegraphics[width=1\textwidth]{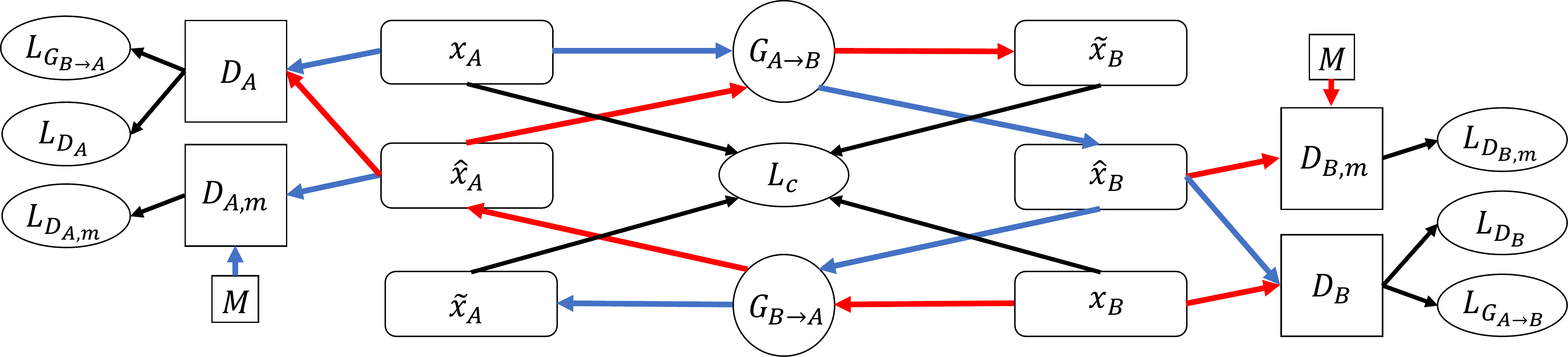} 
\caption{Architecture of our model. The two cycles are shown in blue and red respectively. The black arrows point to the loss functions. We extend the basic CycleGAN architecture with additional discriminators $D_{A,m}$ and $D_{B,m}$.} 
\label{fig:architecture2}
\end{figure*}

Our model is based on Generative Adversarial Networks (GANs)~\cite{goodfellow2014generative}. Vanilla GANs consist of a generator $G$ and a discriminator $D$. The generator tries to generate real looking data from noise, while the discriminator attempts to distinguish the output of the generator from real data. $G$ and $D$ are iteratively trained in a two-player minimax game manner. 
% The goal of the generator is to fool the discriminator by generating data that looks real, i.e., as if it came from the true underlying data generating distribution. 
Since our goal is to transfer music from one domain to another, the generator does not actually get noise as input, but instead real samples from the source domain. In this paper we only deal with translation between two domains at a time, and will hence refer to them as domain $A$ and $B$, where the two domains correspond to music from two different genres. Since the transfer should be symmetric, i.e., we want to transfer samples from $A$ to $B$ and vice versa, our model follows the same structure as the recently introduced CycleGAN~\cite{DBLP:conf/iccv/ZhuPIE17}.
% In this case, the GAN minimax game can formally be written as
% \nb{rewrite minmax game like in original gan paper, i.e., without domains A and B}
% \begin{equation*}
% \min _{\theta _{g}}\max_{\theta _{d}}\left \{ \mathbb{E}_{b}\left [ \log D_{B}\left ( x_{B} \right ) \right ]+\mathbb{E}_{a}\left [ \log \left ( 1-D_{B}\left ( \hat{x}_{B} \right ) \right ) \right ] \right \}
% \end{equation*}
% where $\theta _{g}$ and $\theta _{d}$ are the parameters of the generator $G$ and discriminator $D$ respectively. Once the GAN has converged it can be used to generate new data from noise.
% Our model is based on CycleGAN~\cite{DBLP:conf/iccv/ZhuPIE17} and aims to realize musical style translation between two domains $A$ and $B$. Domains $A$ and $B$ contain music from two different genres, such as Jazz and Classic. 
A CycleGAN basically consists of two GANs that are arranged in a cyclic fashion and trained in unison. One generator transfers data from domain $A$ to $B$ and the other from $B$ to $A$. One discriminator is attached to each generator output.
Figure~\ref{fig:architecture2} shows the architecture of our model. Blue and red arrows denote the domain transfers in the two opposite directions, and black arrows point to the loss functions. $G_{A\rightarrow B}$ and $G_{B\rightarrow A}$ are two generators which transfer data between $A$ and $B$. $D_{A}$ and $D_{B}$ are two discriminators which distinguish if data is real or fake. $D_{A, m}$ and $D_{B, m}$ are two extra discriminators which force the generators to learn more high-level features. Following the blue arrows, $x_{A}$ denotes a real data sample from source domain $A$. $\hat{x}_{B}$ denotes the same data sample after being transferred to target domain $B$, i.e., $\hat{x}_{B}=G_{A\rightarrow B}\left ( x_{A} \right )$. $\tilde{x}_{A}$ denotes the same data sample after being transferred back to the source domain $A$, i.e., $\tilde{x}_{A}=G_{B\rightarrow A}\left ( G_{A\rightarrow B}\left ( x_{A} \right ) \right )$. Equivalently, following the red arrows describes the opposite direction, i.e., the transfer from $B$ to $A$ and back to $B$. $M$ is a dataset containing music from multiple domains, e.g, $M=A\cup B$. $x_{M}$ denotes a data sample from $M$. 
% In the following we will describe the loss functions of our model. 

% CycleGAN is constructed based on two pairs of GAN. Concretely, $A$ and $B$ are two different data domains, and the generators $G_{A\rightarrow B}$ and $G_{B\rightarrow A}$ learn the translation $A\rightarrow B$ and $B\rightarrow A$ simultaneously to fool the discriminators.

% Discriminators $D_{A}$ and $D_{B}$ learn to distinguish if samples are real or fake. 

% \nb{@sumu: I think we should add a small diagram that shows a cyclegan, so we can better explain the cycle consistency loss}

% Thus the total loss function  $L\left ( G_{A\rightarrow B}, G_{B\rightarrow A}, D_{A}, D_{B} \right )$ is   
% \begin{equation*}
% L\left ( G_{A\rightarrow B}, D_{B} \right ) + L\left ( G_{B\rightarrow A}, D_{A} \right ) + \lambda L_{c}\left ( G_{A\rightarrow B}, G_{B\rightarrow A}\right )
% \end{equation*}
% % where $L_{c}\left ( G_{A\rightarrow B}, G_{B\rightarrow A}\right ) = \left \|  G_{B\rightarrow A}\left ( G_{A\rightarrow B}\left ( a \right ) \right ) -a\right \|_{1} + \left \|  G_{A\rightarrow B}\left ( G_{B\rightarrow A}\left ( b \right ) \right ) -b\right \|_{1}$ is the cycle consistency loss, 
% where $\lambda$ is the weight of the cycle loss.  

As in~\cite{DBLP:conf/iccv/ZhuPIE17}, we use the L2 norm for the adversarial loss. For the generators, we have 

\begin{equation*}
L_{G_{A\rightarrow B}} =\left \| D_{B}\left ( \hat{x}_{B} \right ) -1\right \|_{2}
\end{equation*}

\begin{equation*}
L_{G_{B\rightarrow A}}=\left \| D_{A}\left ( \hat{x}_{A} \right ) -1\right \|_{2}
\end{equation*}

To enforce forward-backward consistency, Zhu et al.~\cite{DBLP:conf/iccv/ZhuPIE17} introduce an extra L1 loss term called \emph{cycle consistency loss}: 

\begin{equation*}
L_{c} = \left \|  \tilde{x}_{A} - x_{A}\right \|_{1} + \left \|  \tilde{x}_{B} - x_{B}\right \|_{1}
\end{equation*}

The cycle consistency loss ensures that the input is mapped back to itself after passing it through both generators, i.e., after completing the cycle.  
If the cycle loss is omitted, the generators will suffer from posterior collapse, and there will only be little or no mutual information between the input and output, which is generally undesirable. The cycle consistency loss can also be seen as a regularizer that makes sure the generators do not ignore the input data, but instead retain as much information as necessary to then be able to invert the transformation.

Thus, the total loss function of the generators is 

\begin{equation}\label{eqn:totalgenloss}
L_{G}=L_{G_{A\rightarrow B}}+L_{G_{B\rightarrow A}}+\lambda L_{c}
\end{equation}
Where $\lambda$ is used to weight the contribution of the cycle consistency loss. 
For the standard GAN discriminators, we have

\begin{equation*}
L_{D_{A}}=\frac{1}{2}\left ( \left \| D_{A}\left ( x_{A} \right ) -1\right \|_{2}+\left \| D_{A}\left ( \hat{x}_{A} \right )\right \|_{2} \right )
\end{equation*}

\begin{equation*}
L_{D_{B}}=\frac{1}{2}\left ( \left \| D_{B}\left ( x_{B} \right ) -1\right \|_{2}+\left \| D_{B}\left ( \hat{x}_{B} \right )\right \|_{2} \right )
\end{equation*}

% \begin{equation}
% L_{D}=L_{D_{A}} +L_{D_{B}}
% \end{equation}

% TODO: This stuff should be mentioned in the architecture section

% The discriminators are typical CNNs with convolution layers. 
% The output of $D_{A}$ and $D_{B}$ is close to 1 for the real data and 0 for the fake data (i.e., generated data). The generators can be viewed as \quotes{autoencoders} with ResNet~\cite{DBLP:conf/cvpr/HeZRS16} blocks. They consists of convolution layers, ResNet blocks and transposed convolution layers. We use a sigmoid activation at the output of the last layer which ensures the output is in $\left [ 0, 1 \right ]$. 

% \nb{argument about why we add the two discriminators needs to be worked on more. maybe add an example: for example, jazz music is usually faster and less structured than classic music, but we do not want the generator to simply "scramble" classic music when transferring to jazz, even though  this would fool the basic discriminator. the additional discriminator ensures that the generator still has to generate realistic music, which constrains the generator to retain more of the source song's structure}

% \nb{move some of the explanation from section 6B to here and reduce redundancies}

% One of the most common problems, and the one we also observed in our experiments, is that the discriminator overpowers the generator, i.e., the discriminator becomes too good too fast. The generator then never has a chance to learn, because it gets basically \quotes{nothing right}, and hence does not know how to effectively improve. 

GAN training is highly unstable and the discriminator and generator training needs to be carefully balanced. A common failure mode is when the discriminator is too powerful and overpowers the generator early in training, which results in convergence to a bad local optima. In our setting there is another difficulty: Since the generators need to learn a transformation from a source to a target music genre, they effectively need to learn features of both genres, such that the discriminator can be fooled. It is likely that music genres have a few very distinctive patterns, and that the generators could then simply generate many of these patterns in an attempt to fool the discriminator. Even though the discriminator might be fooled, the output might not sound realistic anymore.
% For example, Jazz music is generally faster and less structured than Classic music. A generator could now figure out that it simply needs to shorten and \quotes{scramble} all notes in a Classical piece to fool the discriminator. In this case, one might argue that the genre transfer was successful since the discriminator was fooled, but the resulting music probably does not sound like real Jazz. 
% This might work well, and the genre transfer might indeed be successful (since the discriminator was successfully fooled), but the music might now sound unpleasant. 
In order to force the generators to learn better high-level features, we add two extra discriminators. The main difference to the standard discriminators is that they are trained to distinguish fake data and data from multiple domains ($M$), instead of just data from the target domain. This helps regularize the generator to stay on the \quotes{music manifold}, and generate plausible, realistic music. More importantly, it causes the generator to retain much of the input's structure, thereby ensuring that the original piece is still recognizable after the genre transfer.
% The network architectures are the same as $D_{A}$ and $D_{B}$. The difference is that we feed real data which comes from both domains. 
% Due to the inputs not only coming from the target domain, it is more difficult to learn a qualified discriminator. 
The loss for these two extra discriminators $D_{A, m}$ and $D_{B, m}$ is

\begin{equation*}
L_{D_{A, m}}=\frac{1}{2}\left ( \left \| D_{A, m}\left ( x_{M} \right ) -1\right \|_{2}+\left \| D_{A, m}\left ( \hat{x}_{A} \right )\right \|_{2} \right )
\end{equation*}

\begin{equation*}
L_{D_{B, m}}=\frac{1}{2}\left ( \left \| D_{B, m}\left ( x_{M} \right ) -1\right \|_{2}+\left \| D_{B, m}\left ( \hat{x}_{B} \right )\right \|_{2} \right )
\end{equation*}
where $M$ denotes mixed real data from multiple domains (here possibly Jazz, Classic and/or Pop). Thus the total loss for the discriminators is

\begin{equation}\label{eqn:totdiscloss}
L_{D, all}=L_{D}+\gamma \left (  L_{D_{A, m}}+L_{D_{B, m}}\right )
\end{equation}

where $\gamma$ is used to weight the extra discriminator losses. To further stabilize the GAN training we add Gaussian noise $N\left ( 0, \sigma_{D} ^{2} \right )$ to the inputs of the discriminators, similar to~\cite{DBLP:journals/corr/SonderbyCTSH16}. This improves the robustness and generalization performance of the model. The effects of adding the extra discriminators and applying noise to the input of all discriminators are evaluated in Sections~\ref{subsec:discinputnoise} and \ref{subsec:styletransferresults}.

% This should be mentioned later in the evaluation
% Also, we add Gaussian noise $N\left ( 0, \sigma_{c} ^{2} \right )$ to the inputs for the sake of generalization. 

\section{Dataset and Preprocessing}\label{sec:dataset}

We train our models on music in the MIDI
% ~\cite{midistandard} 
format, which is a symbolic music representation that resembles sheet music. MIDI (Musical Instrument Digital Interface) was originally created as a standard communication interface between electrical instruments, computers and other devices. Thus, MIDI files do not contain actual sound like MP3 files, but instead so-called MIDI messages. For us, the most relevant are the \emph{Note On} and \emph{Note Off} messages. The \emph{Note On} message indicates that a note is beginning to be played, and it also specifies the velocity (loudness) of that note. The \emph{Note Off} message denotes the end of a note. Each note also has a specified pitch, which in MIDI can range between 0 and 127, corresponding to a note range of $C_{-1}$ to $G_{9}$. A standard piano can play MIDI notes 21 to 108, or equivalently $A_0$ to $C_8$. Velocity values also range between 0 and 127. 
Since MIDI files do not contain any sounds themselves, a MIDI synthesizer is required to actually play them. Such synthesizers can either be hardware devices or pieces of software. The final sound will depend on the implementation of the instrument sounds within the used synthesizer.

To input MIDI files to a neural network, they must first be converted into a matrix, the so-called \emph{piano roll} representation, which can be obtained using the \textsl{pretty\_midi}~\cite{data2014intuitive} and \textsl{Pypianoroll}~\cite{DBLP:conf/aaai/DongHYY18} Python packages. 
% The de-facto standard for representing MIDI music for training neural networks is the so-called \emph{piano roll} representation. 
% In order to transform MIDI files into piano rolls, we use the textsl{pretty\_midi} \cite{data2014intuitive} and \textsl{Pypianoroll}\cite{DBLP:conf/aaai/DongHYY18} Python packages. 
% In general, MIDI files have multiple tracks, where each track can be assigned a different instrument. 
Since MIDI notes can have arbitrary lengths, it is necessary to re-sample the MIDI file in order to discretize time and allow a matrix representation. We use a sampling rate of 16 time steps per \emph{bar}, a common choice in the literature (\hspace{-1sp}\cite{DBLP:conf/ismir/YangCY17,midivaeismir2018}), which means that the shortest possible note is the 16th note. A bar is a segment of time corresponding to a specific number of beats, each of which is represented by a particular note value and the boundaries of the bar are indicated by vertical lines (\emph{bars}) on a music sheet. For example in the common $\frac{4}{4}$ time signature, a bar contains 4 beats and on each beat we can play a quarter note, two eighth notes, four sixteenth notes, and so on. 
Thus, our final piano-roll representation is a $t \times p$ matrix, where $t$ denotes the number of time steps (e.g., $t$=16 for a 1-bar piece), and $p$ denotes the number of pitches. We omit the velocity information by setting all velocities to 127, such that every note has the same loudness. This makes learning easier, since every note can now only be \emph{on} or \emph{off}, instead of taking 128 possible values. Therefore, the piano-roll representation contains a $p$-dimensional $k$-hot vector at each time step, where $k$ is the number of simultaneously played notes. Because notes with the pitch below C1 or above C8 are not very common, we only retain notes between this range, i.e., $p=84$. Thus, the piano-roll for one bar is of size $16\times 84$. Since music has temporal structures we need to consider the relation of consecutive bars. Therefore, similar to MuseGAN~\cite{DBLP:conf/aaai/DongHYY18}, we use phrases consisting of 4 consecutive bars as training samples. Thus, the resulting samples are of size $64\times 84$.

MIDI files can have multiple tracks, where each track can be assigned a different instrument. Since we base our genre transfer solely on note pitches, it is important that we retain as much of the \quotes{content} of the original song as possible. Otherwise, the \quotes{style} of the song might be lost after selecting only a subset of voices. For example, two songs from two different genres with roughly the same melody but different accompaniments would be difficult to classify if we only consider the notes that comprise the melody. 
While some previous works select a limited number of voices to make learning easier (e.g.,\cite{midivaeismir2018}), we simply merge \emph{all} notes of all tracks into a single track. 
By doing so we retain most of the original songs identity, i.e., it is still clearly recognizable as the original song. 
% We find that CNNs seem to have no problem modeling a high number of voices, i.e., simultaneously played notes.
% Thus, a genre classifier can achieve good performance, as we show in our experiments. This is a necessary condition for being able to extract style features and perform style transfer. 
However, since all notes are now played by the same instrument, the music can sound cluttered. We therefore do not use highly complex pieces of music such as symphonies, as the number of different voices and instruments is simply too high. We further omit the drum track, since it often sounds bad when played by another instrument. 

In order to perform domain transfer we require music from different genres. In this paper we use songs from the genres Jazz, Classic and Pop which we collected from various sources. 
% \nb{@sumu: we should make the complete and cleaned dataset available. we will just somehow link it on github. Sumu: At hand I have all the qualified datasets and training datasets after preprocessing which can be used for training directly.}
As the dataset is noisy, we need to perform several preprocessing steps. First, we filter out MIDI files whose first beat does not start at 0. Then we remove songs whose time signature changes throughout the song, or whose time signature is not $\frac{4}{4}$. 
% Second, in some MIDI files, some tracks only play a few notes and are mostly silent otherwise, which introduces data sparsity and impedes the training process. Thus for each MIDI file, we merge (i.e., sum) all the tracks except the drum track (if it has) into a single track, which sounds almost same as the original multi-track one. Most drum tracks have dense distribution of notes. If we merge the drum tracks with other tracks, the merged single track will sound messy, especially for jazz and pop music. 
% Second, we fix the smallest note unit to be a sixteenth note. Because notes with the pitch below C1 or above C8 are not very common, we only retain notes between this range. Thus, a bar for training and testing has shape $16*84$. 
% But we still use all the 128 MIDI notes from C0 to C10 in our symbolic representation. 
After these preprocessing steps, we have a clean dataset consisting of 12,341 Jazz, 16,545 Classic and 20,780 Pop samples, where the length of one sample is equal to one phrase, or four bars. To avoid introducing a bias due to the imbalance of genres, we reduce the amount of samples in the larger dataset to match that of the smaller one. For example, when training on Jazz and Classic, we randomly sample 12,341 phrases from the Classic dataset to match the size of the Jazz dataset.  
% \nb{here, for JC or JP, 12341, for CP, 16545. }
% After these preprocessing steps, we finally get a clean dataset consisting of 11216 phrases for training and 1125 phrases for testing in Jazz domain, 14598 phrases for training and 1947 phrases for testing in Classic domain, 18916 phrases for training and 1864 phrases for testing in Pop domain. 

\section{Architecture Parameters and Training}

% \nb{@Sumu: Write some details here about how you chose the architectures of the generator/discriminator. Did you try architectures from different papers? e.g., midinet, musegan, cyclegan. Did you try different types of CNNs, such as VGG, or omitting the Residual layers? Sumu: I really tried these original models. But midinet and musegan generate music from noise so it cannot meet our requirement. For the CNN without ResNet, I tried several times but I didn't record results. So here I can only say I have tried those original models but all the results are not good. ?}

GAN training is generally unstable, as the generator and discriminator need to be carefully balanced. 
% One of the most common problems, and the one we also observed in our experiments, is that the discriminator overpowers the generator, i.e., the discriminator becomes too good too fast. The generator then never has a chance to learn, because it gets basically \quotes{nothing right}, and hence does not know how to effectively improve. 
Many techniques have been introduced in order to stabilize GAN training~\cite{ganhacks}, of which we employ several, such as using instance normalization~\cite{DBLP:journals/corr/UlyanovVL16} and LeakyReLU~\cite{maas2013rectifier} activations. 
% \nb{explain instance norm and relu/leakyrelu} 
% We implement our models using TensorFlow. 
During development we experimented with different architectures for the generator and discriminator, before settling on those shown in Tables~\ref{table:D} and \ref{table:G}. The inputs to the generators and discriminators have the shape $\left ( batchsize, 64, 84, 1 \right )$. Before feeding the samples to the models, we normalize the pitch values to the range [0,1].
We use the Adam~\cite{DBLP:journals/corr/KingmaW13} optimizer with an initial learning rate of $\alpha=0.0002$. The momentum decay rates are set to $\beta_{1} =0.5$ and $\beta_{2} =0.999$. 
% The learning rate would decay 
% \nb{@sumu what is the exact decay schedule? Sumu: $lr = args.lr * (args.epoch-epoch) / (args.epoch-args.epoch\_step)$, from Cyclegan, they set the epoch\_step=10. }
% after ten epochs to avoid model collapsed into local minima. 
As suggested by \cite{DBLP:conf/iccv/ZhuPIE17}, we set $\lambda=10$ in Equation~\ref{eqn:totalgenloss}. Also, we choose $\gamma=1$ in Equation~\ref{eqn:totdiscloss}. We train each model for a maximum of 30 epochs, or until the cycle loss converges, and set the batch size to $16$. 

\begin{table}[tb]
\centering
\caption{Discriminator architecture}
\label{table:D}
\begin{tabular}{cccccc}
\hline
\multicolumn{6}{l}{Input: $\left ( batch size \times 64\times 84\times 1 \right )$}  \\ \hline
layer    & filter        & stride        & channel   & instance norm   & activation  \\ \hline
conv     & $4\times 4$   & $2\times 2$   & 64        & False           & LReLu       \\
conv     & $4\times 4$   & $2\times 2$   & 256       & True            & LReLu       \\
conv     & $1\times 1$   & $1\times 1$   & 1         & False           & None        \\ \hline
\multicolumn{6}{l}{Output: $\left ( batch size \times 16\times 21\times 1 \right )$} \\ \hline
\end{tabular}
\end{table}

\begin{table}[tb]
\centering
\tabcolsep=0.12cm
\caption{Generator architecture}
\label{table:G}
\begin{tabular}{cllccc}
\hline
\multicolumn{6}{l}{Input: $\left ( batch size \times 64\times 84\times 1 \right )$} \\ \hline
layer & \multicolumn{1}{c}{filter} & \multicolumn{1}{c}{stride} & channel & instance norm & activation \\ \hline
conv & \multicolumn{1}{c}{$7\times 7$} & \multicolumn{1}{c}{$1\times 1$} & 64 & True & ReLu \\
conv & \multicolumn{1}{c}{$3\times 3$} & \multicolumn{1}{c}{$2\times 2$} & 128 & True & ReLu \\
conv & \multicolumn{1}{c}{$3\times 3$} & \multicolumn{1}{c}{$2\times 2$} & 256 & True & ReLu \\ \hline
\multirow{2}{*}{$10 \times $ ResNet} & $3\times 3$ & $1\times 1$ & 256 & True & ReLu \\
 & $3\times 3$ & $1\times 1$ & 256 & True & ReLu \\ \hline
deconv & $3\times 3$ & $2\times 2$ & 128 & True & ReLu \\
deconv & $3\times 3$ & $2\times 2$ & 64 & True & ReLu \\
deconv & $7\times 7$ & $1\times 1$ & 1 & False & Sigmoid \\ \hline
\multicolumn{6}{l}{Output: $\left ( batch size \times 64\times 84\times 1 \right )$} \\ \hline
\end{tabular}
\end{table}

\begin{table}[tb]
\centering
\caption{Classifier architecture}
\label{table:C}
\begin{tabular}{cccccc}
\hline
\multicolumn{6}{l}{Input: $\left ( batch size \times 64\times 84\times 1 \right )$} \\ \hline
layer & filter & stride & channel & instance norm & activation \\ \hline
conv & $1\times 12$ & $1\times 12$ & 64 & False & LReLu \\
conv & $4\times 1$ & $4\times 1$ & 128 & True & LReLu \\
conv & $2\times 1$ & $2\times 1$ & 256 & True & LReLu \\
conv & $8\times 1$ & $8\times 1$ & 512 & True & LReLu \\
conv & $1\times 7$ & $1\times 7$ & 2 & False & Softmax \\ \hline
\multicolumn{6}{l}{Output: $\left ( batch size \times 2 \right )$} \\ \hline
\end{tabular}
\end{table}

% The architecture of the style classifier is similar to the discriminators. We used five convolution layers: the first layer uses filters of shape $\left (1, 12 \right )$ and strides of $\left (1, 12 \right )$, the second layer uses filters of shape $\left (4, 1 \right )$ and strides of $\left (4, 1 \right )$, the third layer uses filters of shape $\left (2, 1 \right )$ and strides of $\left (2, 1 \right )$, the fourth layer uses filters of shape $\left (8, 1 \right )$ and strides of $\left (8, 1 \right )$ and the last layer uses filters of shape $\left (1, 7 \right )$ and strides of $\left (1, 7 \right )$. We applied the instance norm to the layer 2, 3, 4 and leaky ReLu activation to layer 1, 2, 3, 4. 

\section{Experimental Results}

Evaluating the performance of a music generation system is difficult since the goodness of music is a highly subjective measure. Evaluating style and domain transfer is slightly simpler, because one effectively generates aligned pairs of samples, e.g., $x_A$ and $\hat{x}_B$, where both samples have a domain label. Thus, we can train a style classifier $C_{A,B}$ to distinguish between genre A and B, and then apply it to $x_A$ and $\hat{x}_B$. If the style transfer works, $C_{A,B}$ will classify $x_A$ as $A$, and $\hat{x}_B$ as $B$. The more confident the classifier is, the stronger the genre transfer. In the following we describe the style classifier $C_{A,B}$, before describing how the GAN training can be improved by applying Gaussian noise to the discriminator inputs. Finally, we evaluate the style transfer effectiveness of multiple models.

% In this section, we first trained classifiers based on real data with different $\sigma_{c}$ in order to check the robustness of the classifiers. Second, we did value search for the three models trained on jazz and classic domain in order to choose an proper $\sigma_{d}$. Third, we trained three different models with best chosen $\sigma_{d}$ and compared the training process and average accuracy such that we chose a best model. Finally, we experimented on other domain pairs like jazz to pop, classic to pop. 

\subsection{Genre Classifier}\label{subsec:valstyleclass}

\begin{table}[t]
\centering
\tabcolsep=0.085cm
\caption{Average genre classifier accuracy with Gaussian input noise added during testing to evaluate robustness.}
\label{table:1}
\begin{tabular}{c|cccccc}
$\sigma_{C}$     & 0       & 0.01    & 0.1     & 0.2      & 0.3      & 0.5\\ \hline
Jazz vs. Classic & 88.89\% & 88.53\% & 87.87\%  & 84.71\%  & 83.07\%  & 74.93\%\\
Classic vs. Pop  & 84.66\% & 83.42\% & 81.97\%  & 81.12\%  & 78.14\%  & 70.28\%\\
Jazz vs. Pop     & 67.07\% & 66.18\% & 63.78\%  & 62.40\%  & 61.72\%  & 59.96\%
\end{tabular}
\end{table}

To evaluate whether our model really learns the translation among different genres, we build a binary classifier $C_{A,B}$ that outputs a probability distribution over domains A and B. The architecture of the genre classifier is shown in Table~\ref{table:C}. We apply a softmax activation to the two output neurons of the last layer, and optimize the classifier with a cross-entropy loss. We train the genre classifiers on real data from two domains, e.g., Jazz and Classic. The data is the same as that used during the GAN training, and we use a 90/10 train/test split. The performance of the genre classifiers on the test sets is shown in the first column of Table~\ref{table:1}. The accuracy on Jazz vs. Classic and Classic vs. Pop is quite high, with 88.89\% and 84.66\% respectively. The classifier's performance on Jazz vs. Pop is significantly lower, indicating that the two genres are more similar, at least when only considering note pitches. 

The genre classifiers are trained only on real data. However, we want to use them to evaluate whether a domain transfer was successful. For this, we need to apply the classifier on data that has been passed through a generator. If the generator successfully recovered the underlying data generating distribution, the two cases are the same. However, in practice we have to assume that the generator is not perfect and that the generated data that is somehow different from real data. Therefore, the train and test set for the genre classifier effectively come from slightly different distributions, i.e., we are applying the genre classifier on fake data, where during training it has only ever seen real data. This could potentially negatively affect the usefulness of our genre classifier, as we do not know how well it will generalize to fake data. In this paper we investigate the robustness of our genre classifier in the case where the generators apply Gaussian noise to the inputs. This is of course a simplification, since the true transformations applied by the generators are more complex. Table~\ref{table:1} shows how the performance of the style classifier changes when Gaussian noise ($\mathcal{N}(0,\sigma_C^2)$) is applied to the inputs of the classifier. Note that the inputs (note pitches) are normalized to lie in the range [-1,1]. The results show that the genre classifier is robust even when adding noise that is large relative to the input value range (i.e. $\mathcal{N}(0,0.5^2)$). We therefore conclude that the genre classifier learned salient features and cannot easily be broken by random noise. We leave the evaluation of the classifier's robustness against more sophisticated, possibly even adversarial, noise for future work.

In the following we will use the genre classifier $C_{A,B}$ to evaluate the results of the domain transfer. When considering a transfer from $A$ to $B$, $C_{A,B}$ reports the probability $P_A(x)$ if the source genre is $A$, and $P_B(x)$ if the source genre is $B$. We consider a domain transfer from $A$ to $B$ as successful if $P_A(x_A)=C_{A,B}(x_A)>0.5$ AND $P_A(\hat{x}_B)=C_{A,B}(\hat{x}_B)<0.5$. In other words: If the source style is considered to be more likely before the transfer, and less likely after the transfer. Among the successful domain transfers, we define the \emph{strength} of the domain transfer in one direction (A $\rightarrow$ B $\rightarrow$ A) as 
% \nb{the second term: without hat on x?}
\begin{equation*}
S_{A\rightarrow B}^D=\frac{P(A|x_A)-P(A|\hat{x}_B)+P(A|\tilde{x}_A)-P(A|\hat{x}_B)}{2} 
\end{equation*} 
For the other direction i.e., (B $\rightarrow$ A $\rightarrow$ B), $S_{B\rightarrow A}^D$ is defined analogously. The final domain transfer strength of a particular model is defined as the average of the strengths in both directions

\begin{equation*}
S_{tot}^D=\frac{1}{2}(S_{B\rightarrow A}^D+S_{A\rightarrow B}^D)
\end{equation*}

% \begin{equation*}
% S_D=\frac{P(A|x_A)-P(A|\hat{x}_B)+P(B|x_B)-P(B|\hat{x}_A)}{2} 
% \end{equation*}

The maximum strength that can be achieved is $S_{tot}^D=1$ if for both directions, the source style's probability is equal to 1 before the transfer, equal to 0 after the transfer, and again equal to 1 after completing the cycle. 
For the remainder of this paper we will use this metric to determine how well a model can perform domain transfer. However, a model that does not retain any structure of the original input can still achieve $S_{tot}^D=1$ if the generators learn to perfectly invert each other. Therefore, human judgment is still necessary to determine whether a model performs well. Generally, we are looking for a model that transforms a piece of music from a source to a target genre while retaining as much of the source's content as possible.   

% \nb{@Sumu: Can you include more values for $\sigma_c$? Basically until it starts to deteriorate strongly, e.g, at 5 or 10 Sumu: I'll train them immediately. }
% \nb{@Sumu: how exactly is the accuracy computed? Sumu: for example, jazz and classic, I label jazz music as [1, 0], classic music as [0, 1], then use softmax on the 2 output neurons [x1, x2]. x1 is bigger then it is jazz otherwise classic. }
% \begin{table}[tbh]
% \centering
% \caption{Average Accuracy of classifiers}
% \label{table:1}
% \begin{tabular}{c|ccccc}
% $\sigma_{C}$     & 0       & 0.01    & 0.1     & 1       & 2       \\ \hline
% Jazz vs. Classic & 88.89\% & 88.53\% & 89.73\% & 90.31\% & 87.69\% \\
% Classic vs. Pop  & 84.66\% & 84.76\% & 85.57\% & 87.55\% & 82.73\% \\
% Jazz vs. Pop     & 67.07\% & 66.98\% & 67.64\% & 66.58\% & 68.00\%
% \end{tabular}
% \end{table}

\begin{table}[t]
\centering
\caption{Losses of the \emph{base} model after 20 epochs. Only for $\sigma_D=1$ did the cycle loss converge to 0 and the discriminator and generator are in balance.}
\label{table:2}
\begin{tabular}{c|cccccc}
$\sigma_{D}$ & 0    & 0.01 & 0.1  & \textbf{1}    & 3    & 5    \\ \hline
$L_{c}$ & 0.37 & 0.98 & 0.20 & \textbf{0.00} & 0.29 & 0.87 \\
$L_{G}$ & 1.20 & 1.87 & 1.00 & \textbf{0.52} & 0.80 & 1.56 \\
$L_{D}$ & 0.36 & 0.27 & 0.41 & \textbf{0.49} & 0.50 & 0.44
\end{tabular}
\end{table}

\subsection{Discriminator Input Noise to Stabilize GAN Training} \label{subsec:discinputnoise}

% \nb{this part needs to be adapted with the complete hyper parameters for sigma\_D now that sumu did a new search for the CP model}

% \nb{now we did a full search on sigma\_D for base, partial and full. thus, for each domain-pair we train 18 models and pick the best sigma\_D for base, partial and full respectively. Thus, we should merge this section with "style transfer"}

% GAN training is generally unstable, as the generator and discriminator need to be carefully balanced. One of the most common problems, and the one we also observed in our experiments, is that the discriminator overpowers the generator, i.e., the discriminator becomes too good too fast. The generator then never has a chance to learn, because it gets basically \quotes{nothing right}, and hence does not know how to effectively improve. Many techniques have been introduced in order to stabilize GAN training~\cite{ganhacks}, of which we employed several, such as using instance normalization and LeakyReLU activations. 

% In this section we discuss a method for stabilizing GAN training and how it influences the domain transfer results. 
In order to force the generators and discriminators to learn better features, i.e., avoid overfitting on spurious patterns, and hence improve generalization, we add Gaussian noise $\mathcal{N}(0, \sigma_{D})$ to both real and fake inputs of the discriminators, similar to \cite{DBLP:journals/corr/SonderbyCTSH16}. We train models for each domain pair with 6 different values for $\sigma_D$. For each domain pair and $\sigma_D$ value we train three different models: A \emph{base} model without extra discriminators, a \emph{partial} model with $D_{A, m}$ and $D_{B, m}$ where $m \in M=A\cup B$ and a \emph{full} model with $D_{A, m}$ and $D_{B, m}$ where $m\in M=A\cup B\cup C$. Since we have three genres in total, $C$ is always the remaining genre on which none of the base discriminators is trained. For simplicity, we henceforth refer to the three models as $M_{base}$, $M_{partial}$ and $M_{full}$. This results in a total of $3*3*6=54$ models, from which we pick the best ones according to our domain transfer strength metric $S^D_{tot}$. 
% ... \nb{update with final values and somehow clarify base models vs models with additional discriminators}. 
% Note that in this step we only train \emph{base} models, i.e., models without the additional discriminators $D_{A,m}$ and $D_{B,m}$. 
For the sake of brevity, we only show the hyper parameter search results for the \emph{base} model trained on the Jazz and Classic domains. Table~\ref{table:2} shows the effect of different values for $\sigma_D$ on the cycle consistency loss ($L_c$), generator loss ($L_G$) and discriminator loss ($L_D$), respectively. For $\sigma_{D}=1$ the cycle loss converges to zero, and the discriminator and generator losses are balanced, which is generally an indicator that the model converged to a good optima and did not experience a failure mode. Table~\ref{table:3} shows the style transfer performance of the same model. According to our genre transfer evaluation metric, the model with $\sigma_{D}=1$ performs best ($S^D_{tot}=69.7\%$), which is consistent with the results from Table~\ref{table:2}. 
% Table~\ref{table:3} further shows that the style transfer works in almost all cases. Choosing different values for $\sigma_{D}$ changes the success rate of the style transfer in an asymmetric way. 
% It is still not clear which model is the best, since a high style transfer success rate does not tell us anything about whether the content of the original sample has been preserved enough. This is a subjective decision which can currently only be made by listening to the generated samples. We will discuss this further in the next section. 
% For now, we choose one value of $\sigma_D$ for each domain pair according our style transfer evaluation metric and the results presented in Table~\ref{table:3}. 
We found that in order to find a model with good performance on a particular domain pair, it is necessary to perform a new parameter search over different values of $\sigma_D$.

\begin{table}[t]
\centering
\tabcolsep=0.085cm
\caption{Genre transfer performance of the \emph{base} model measured by a genre classifier with $\sigma_{C}=1$. A contains Jazz pieces and B contains Classic pieces. 
% The percentages of A, A2B, A2B2A denote the average accuracy of the phrases which are classified as jazz. The percentages of B, B2A, B2A2A denote the average accuracy of the phrases which are classified as classic. 
}
\label{table:3}
\begin{tabular}{c|cccccc}
$\sigma_{D}$ & 0       & 0.01    & 0.1     & \textbf{1}       & 3       & 5       \\ \hline
A            & 88.09\% & 88.09\% & 88.09\% & 88.09\% & 88.09\% & 88.09\% \\
A$\rightarrow$B          & 31.38\% & 5.82\% & 29.16\% & 20.62\% & 12.18\% & 19.47\% \\
A$\rightarrow$B$\rightarrow$A        & 84.71\% & 99.82\% & 84.36\% & 88.18\% & 87.91\% & 34.13\% \\
\hline
B            & 92.53\% & 92.53\% & 92.53\% & 92.53\% & 92.53\% & 92.53\% \\
B$\rightarrow$A          & 48.80\% & 56.26\% & 31.20\% & 20.71\% & 61.78\% & 90.67\% \\
B$\rightarrow$A$\rightarrow$B        & 89.24\% & 89.87\% & 89.33\% & 92.53\% & 90.67\% & 90.67\%\\ \hline
$S_{tot}^D$ & 48.5\% & 61.5\%&58.4\% &\textbf{69.7}\% & 52.8\%&20.5\%
\end{tabular}
\end{table}

\subsection{Genre Transfer}\label{subsec:styletransferresults}
\begin{table}[t]
\centering
\caption{Genre transfer performance of three different models trained on Jazz and Classic. A: Jazz, B: Classic}
\label{tab:jc_final_styletransfer_performance}
\begin{tabular}{c|ccc}
 & \begin{tabular}[c]{@{}c@{}}$M_{base}$\\ $\sigma_{D}=1$\end{tabular} & \begin{tabular}[c]{@{}c@{}}$M_{partial}$\\ $\sigma_{D}=0$\end{tabular} & \begin{tabular}[c]{@{}c@{}}$M_{full}$\\ $\sigma_{D}=0.01$\end{tabular} \\ \hline
A & 88.09\% & 88.09\% & 88.09\% \\
A$\rightarrow$B & 20.62\% & 9.87\% & 20.00\% \\
A$\rightarrow$B$\rightarrow$A & 88.18\% & 87.73\% & 85.16\% \\
\hline
B & 92.53\% & 92.53\% & 92.53\% \\
B$\rightarrow$A & 20.71\% & 25.87\% & 20.09\% \\
B$\rightarrow$A$\rightarrow$B & 92.53\% & 89.51\% & 90.49\% \\ \hline
$S_{tot}^D$ & 69.7\% & 71.6\% & 69.0\%
\end{tabular}
\end{table}

% \nb{mention how the accuracies in the tables for style transfer are to be interpreted. also mention cases for when the style transfer worked well, i.e, with percentages}

% \nb{give some more insight here into how the genre transfer works. i.e., talk about when/why it works/doesnt work. say sth about jazz and pop are quite hard to distinguish, also for humans, possibly due to the instrumentation and velocity information being missing. }

In this section we evaluate the genre transfer performance of our final models. For each model we indicate the specific value for $\sigma_D$ that was used during training. We present domain transfer results on three different domain pairs: Jazz and Classic, Classic and Pop, and Jazz and Pop. For each of these domain pairs we show the results of the models $M_{base}$, $M_{partial}$ and $M_{full}$.
% three different models: A \emph{base} model without extra discriminators, a \emph{partial} model with $D_{A, m}$ and $D_{B, m}$ where $m \in M=A\cup B$ and a \emph{full} model with $D_{A, m}$ and $D_{B, m}$ where $m\in M=A\cup B\cup C$. Where $C$ is the remaining genre on which none of the base discriminators is trained. For simplicity, we henceforth refer to the three models as $M_{base}$, $M_{partial}$ and $M_{full}$. 
Tables~\ref{tab:jc_final_styletransfer_performance}, \ref{tab:cp_final_styletransfer_performance} and \ref{tab:jp_final_styletransfer_performance} show the average genre transfer results of our final models. The tables show the probabilities that the genre classifier $C_{A,B}$ assigned to the source genres. In all cases the transfer is successful, i.e., the genre classifier assigns high probability to the source genre before the transfer, and low probability after the transfer. Please note that since our classifier is binary, a low probability of the source genre is equivalent to a high probability of the target genre. 
For most models, especially for the Jazz/Classic and Classic/Pop pairs, the genre transfer is very strong, as can be seen from the high values of $S^D_{tot}$. For Jazz and Pop, the genre transfer is still successful, but less strong on average. This is due to the fact that the genre classifier $C_{A,B}$ cannot distinguish Pop and Jazz as easily as the other genre pairs. 
% Using a more feature-rich input representation by, e.g., including velocity and instrumentation should help make the two genre more easily separable and thereby improve the genre transfer performance.

\begin{figure}[t] 
\centering
\includegraphics[width=0.49\textwidth, height=0.64\columnwidth]{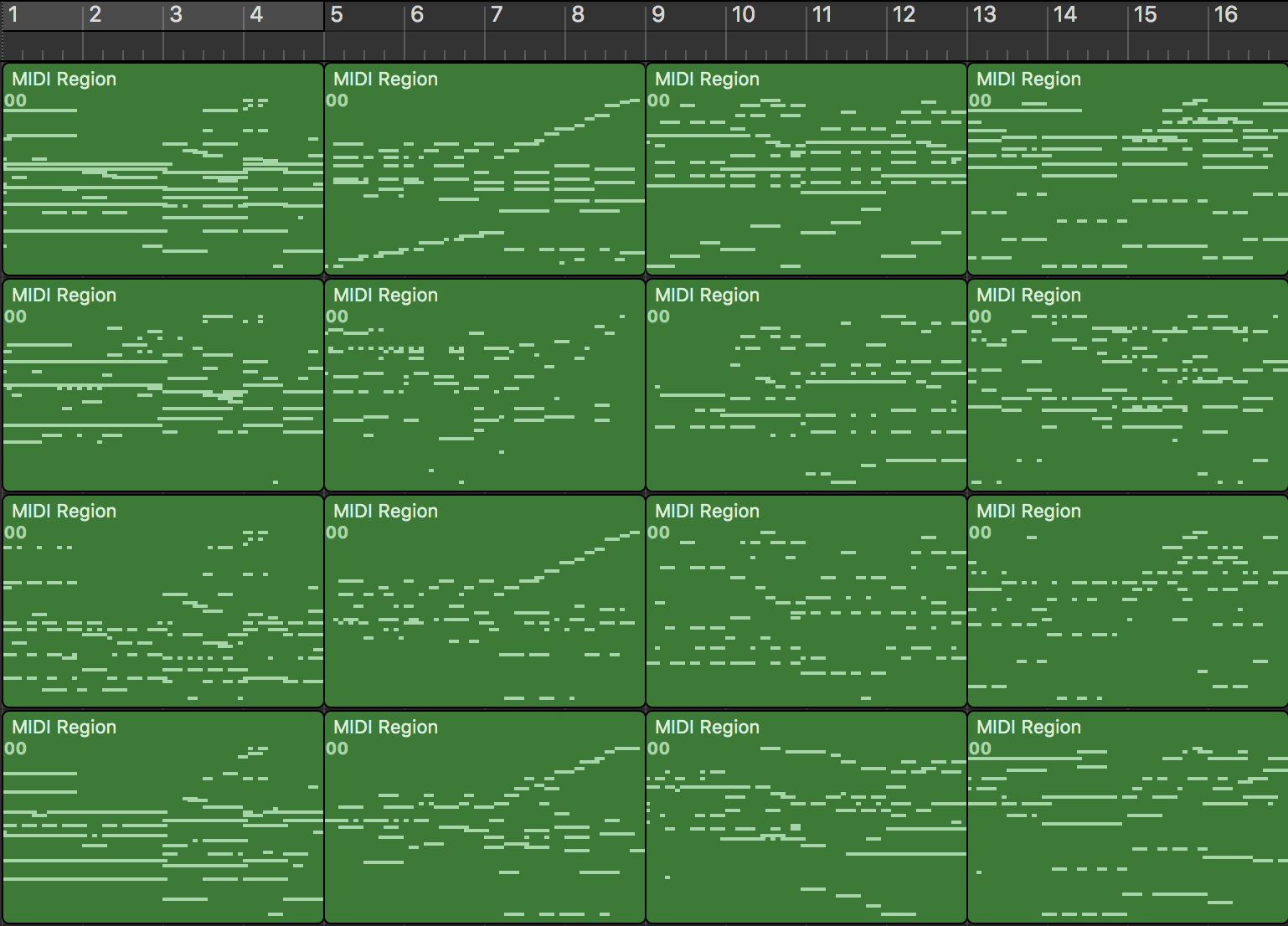} 
\caption{Samples transferred from Jazz to Classic. The first row contains original samples from the test set of domain A (Jazz). The remaining rows show the results from the domain transfer done by models $M_{base}$, $M_{partial}$ and $M_{full}$ respectively (see Table~\ref{tab:jc_final_styletransfer_performance} for more details on the models).} 
\label{fig:comparison AtoB}
\end{figure}

\begin{figure}[t] 
\centering
\includegraphics[width=0.49\textwidth, height=0.64\columnwidth]{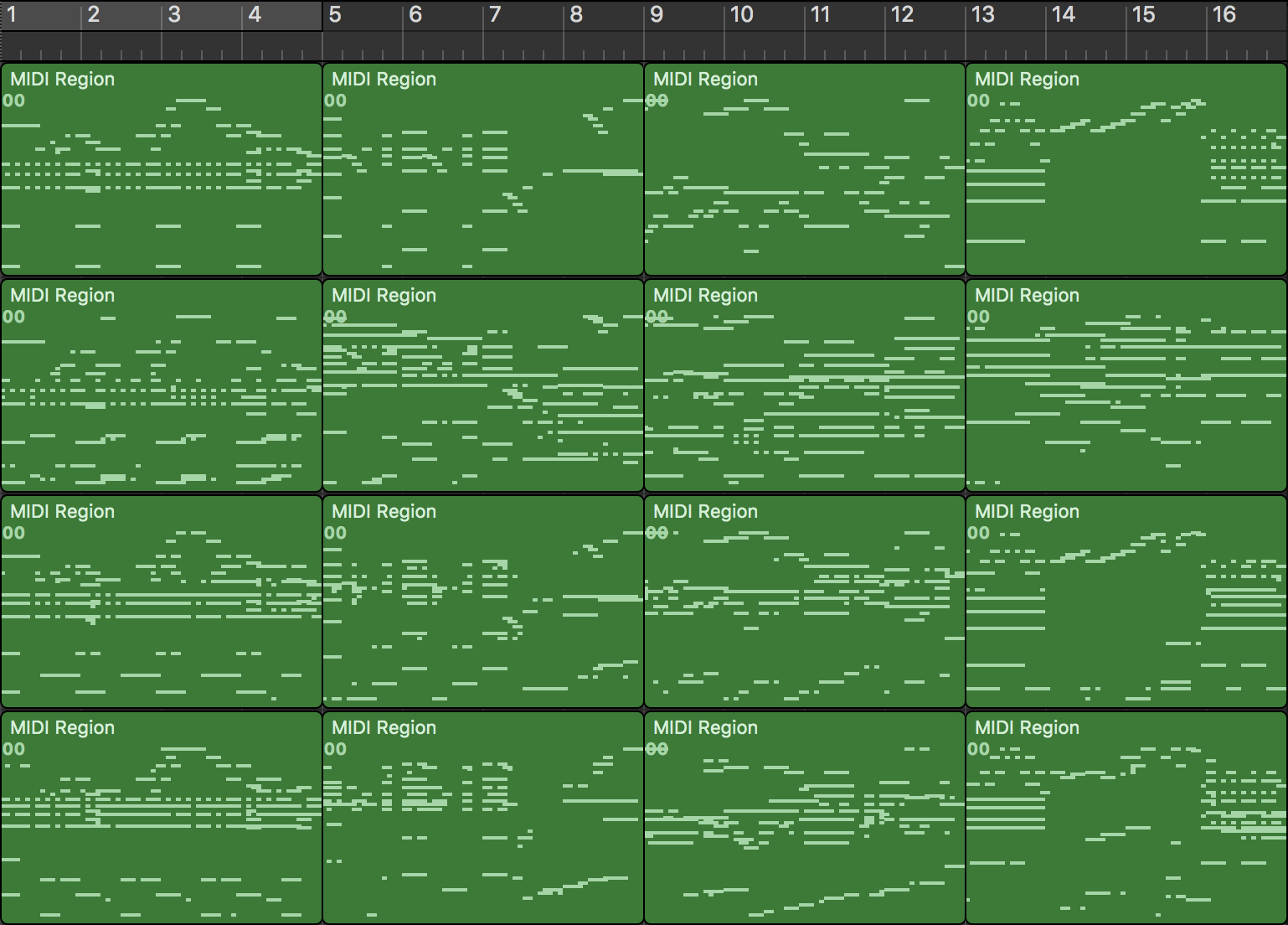} 
\caption{Samples transferred from Classic to Jazz. The first row contains original samples from the test set of domain B (Classic). The remaining rows show the results from the domain transfer done by models $M_{base}$, $M_{partial}$ and $M_{full}$ respectively (see Table~\ref{tab:jc_final_styletransfer_performance} for more details on the models).} 
\label{fig:comparison BtoA}
\end{figure}

At first glance, adding the additional discriminators in $M_{partial}$ and $M_{full}$ does not have a clear benefit, at least measured by our domain transfer metric $S^D_{tot}$. However, as explained in Section~\ref{sec:architecture}, adding the additional discriminators encourages the generators to stay on the \quotes{music manifold} and has the effect of retaining more of the source's structure. This can be seen from Figures~\ref{fig:comparison AtoB} and \ref{fig:comparison BtoA}, where we show some examples for both transfer directions of the Jazz and Classic domain pair. The first row contains four original pieces from the test set. The second row corresponds to the outputs of $M_{base}$, the third row to $M_{partial}$ and the last row to $M_{full}$. $M_{base}$ changes the input too much in most cases, especially when transferring from Classic to Jazz (Figure~\ref{fig:comparison BtoA}). The differences between $M_{partial}$ and $M_{full}$ are less pronounced, but we find that $M_{full}$ consistently produces better results, i.e., a clearly audible domain transfer while leaving the original melody largely intact. This final evaluation is subjective and we leave the development of better genre transfer metrics for future work. Overall, the genre transfer from Jazz to Classic seems to be most noticeable. Generally, the original songs sound better than the transferred once, indicating that the GAN training needs further improvement to produce better sounding music.

% Then, we tested the accuracy of three models using the same classifier with $\sigma_{C}=1$ as above and the trade-off parameters $\lambda=10, \gamma=1$. 

% \nb{fix this sentence}
% Similarly, we also did value search for $M_{partial}$ and $M_{full}$ and finally fixed $\sigma_{D}$ for three models. $\sigma_{D}=1$ for $M_{base}$, $\sigma_{D}=0$ for $M_{partial}$ and $\sigma_{D}=0.01$ for $M_{full}$. 

% \nb{@gino @sumu: Decide on final midi images. maybe the best model we now chose + two other interesting ones, for example one that changes the source melody too much, and one that does not do enough transfer, i.e., keeps the source melody too much 
% }

% \nb{This table is for A=J and B=C. We will train 6 more models, which will result in two more tables, one for JP and one for CP. These will be the final results of our paper. For each of those models we will select some pieces and create midi files (including pictures from garageband) and mp3 files. We will show the best results like in Fig2 and Fig3}

\begin{table}[t]
\centering
\caption{Genre transfer performance of three different models trained on Classic and Pop. A: Classic, B: Pop}
\label{tab:cp_final_styletransfer_performance}
\begin{tabular}{c|ccc}
 & \begin{tabular}[c]{@{}c@{}}$M_{base}$\\ $\sigma_{D}=0.1$\end{tabular} & \begin{tabular}[c]{@{}c@{}}$M_{partial}$\\ $\sigma_{D}=1$\end{tabular} & \begin{tabular}[c]{@{}c@{}}$M_{full}$\\ $\sigma_{D}=1$\end{tabular} \\ \hline
A & 86.91\% & 86.91\% & 86.91\% \\
A$\rightarrow$B & 45.12\% & 38.57\% & 26.61\% \\
A$\rightarrow$B$\rightarrow$A & 83.26\% & 87.39\% & 87.18\% \\
\hline
B & 80.04\% & 80.04\% & 80.04\% \\
B$\rightarrow$A & 49.14\% & 23.13\% & 24.30\% \\
B$\rightarrow$A$\rightarrow$B & 72.48\% & 79.45\% & 80.15\%\\ \hline
$S_{tot}^D$ & 33.6\% & 52.6\%& 58.1\%
\end{tabular}
\end{table}

\begin{table}[h!]
\centering
\caption{Genre transfer performance of three different models trained on Jazz and Pop. A: Jazz, B: Pop}
\label{tab:jp_final_styletransfer_performance}
\begin{tabular}{c|ccc}
 & \begin{tabular}[c]{@{}c@{}}$M_{base}$\\ $\sigma_{D}=0.01$\end{tabular} & \begin{tabular}[c]{@{}c@{}}$M_{partial}$\\ $\sigma_{D}=0.01$\end{tabular} & \begin{tabular}[c]{@{}c@{}}$M_{full}$\\ $\sigma_{D}=0$\end{tabular} \\ \hline
A & 60.53\% & 60.53\% & 60.53\% \\
A$\rightarrow$B & 21.60\% & 14.84\% & 23.64\% \\
A$\rightarrow$B$\rightarrow$A & 57.51\% & 58.84\% & 60.53\% \\
\hline
B & 73.60\% & 73.60\% & 73.60\% \\
B$\rightarrow$A & 40.62\% & 43.11\% & 49.24\% \\
B$\rightarrow$A$\rightarrow$B & 74.40\% & 72.09\% & 73.06\%\\ \hline
$S_{tot}^D$ & 35.4\%& 37.3\% & 30.5\%
\end{tabular}
\end{table}

We believe that the results presented in this paper show that GAN-based genre and style transfer for music is a promising direction. In the future we plan to incorporate instrumentation as well as note durations and velocities. Adding these factors, especially instrumentation, should make the genre transfers more easily audible for humans. 
% Including velocities and instrumentation could further improve the results, which we leave to future work. 
% In MIDI-VAE, the style transfer is mostly achieved through change of instrumentation, and to a lesser extent by adapting the note pitches and velocities. 
To complement the presented evaluation, we provide audio samples corresponding to Figures~\ref{fig:comparison AtoB} and \ref{fig:comparison BtoA}, as well as a few samples of famous songs.\footnote{Audio samples: \\ \url{www.youtube.com/channel/UCs-bI_NP7PrQaMV1AJ4A3HQ}}
% \footnote{\url{https://goo.gl/hWyNGB}}

% From Table 5, the second model achieved the best average accuracy. However, by listening to the generated samples, we found that actually, the third model learned a better generators which could keep the original melody. In Fig.1 and Fig.2, the first piano-rolls are the origin, followed by the transfer of base model, model with extra D on JC and model with extra D on JCP. Obviously, the third model almost retained the melody of the origin. 

% However, the generated samples from the third model keep too many features from the origin that we can hardly distinguish the style translation. This may be caused by the large contribution of the extra discriminators. Based on the third model, we again did value search work on $\gamma$ to explore how these extra discriminators $D_{A, m}$ and $D_{B, m}$ can influence the model learning. 

\section{Conclusion}
In this paper we present, to the best of our knowledge, the first application of GANs to symbolic music domain transfer. We extend the standard CycleGAN model with additional discriminators to regularize the generators. We show that these discriminators improve the generated music by encouraging the generators to preserve the structure of the input, while still performing strong domain transfer. The genre transfer cannot only be picked up by a neural network classifier, but can be heard by the untrained ear. Furthermore, the resulting music has complex structure and generally sounds harmonic. 
In the future we plan to develop more objective genre transfer metrics, and further investigate the generalization capabilities and robustness of the genre classifier metric. Incorporating richer features such as velocities, note durations and instrumentation could further improve the results and make genre transfers more convincing and realistic. While in this paper we mainly focused on evaluating the basic CycleGAN architecture, more sophisticated architectures should be explored as well. 

% \cleardoublepage
% \pagebreak

\bibliographystyle{IEEEtran}
\bibliography{IEEEabrv,references}

\end{document}